# Automatic Performance Debugging of SPMD Parallel Programs

Xu Liu

Institute of Computing Technology
Chinese Academy of Sciences
Beijing 100190, China
xu.liu@rice.edu

Lin Yuan

Institute of Computing Technology
Chinese Academy of Sciences
Beijing 100190, China
yuanlin@ncic.ac.cn

Jianfeng Zhan

Institute of Computing Technology
Chinese Academy of Sciences
Beijing 100190, China
jfzhan@ncic.ac.cn

Bibo Tu

Institute of Computing Technology
Chinese Academy of Sciences
Beijing 100190, China

Dan Meng

Institute of Computing Technology
Chinese Academy of Sciences
Beijing 100190, China

## Abstract

Automatic performance debugging of parallel applications usually involves two steps: automatic detection of performance bottlenecks and uncovering their root causes for performance optimization. Previous work fails to resolve this challenging issue in several ways: first, several previous efforts automate analysis processes, but present the results in a confined way that only identifies performance problems *with apriori knowledge*; second, several tools take exploratory or confirmatory data analysis to automatically discover relevant performance data relationships. However, these efforts do not focus on locating performance bottlenecks or uncovering their root causes.

In this paper, we design and implement an innovative system, *AutoAnalyzer*, to automatically debug the performance problems of single program multi-data (SPMD) parallel programs. Our system is unique in terms of two dimensions: first, *without any apriori knowledge*, we automatically locate bottlenecks and uncover their root causes for performance optimization; second, our method is *lightweight in terms of size of collected and analyzed performance data*. Our contribution is three-fold. First, we propose a set of simple performance metrics to represent behavior of different processes of parallel programs, and present two effective clustering and searching algorithms to locate bottlenecks. Second, we propose to use the rough set algorithm to automatically uncover the root causes of bottlenecks. Third, we design and implement the *AutoAnalyzer* system, and use two production applications to verify the effectiveness and correctness of our methods. According to the analysis results of *AutoAnalyzer*, we optimize two parallel programs with performance improvements by minimally 20% and maximally 170%.

*Keywords*  Parallel programs; automatic performance diagnose;

root causes of bottlenecks; performance optimization

## 1. Introduction

Generally, parallel programs, like bio-informatics or seismic applications, are resource-intensive and time-consuming. How to utilize the limited resources efficiently is a challenging issue for parallel programmer, especially non-experts without deep knowledge of computer science. So it is a crucial task to develop an automatic performance debugging tool to help application programmers *analyze the programs' behavior*, *locate performance bottlenecks* (in short *bottlenecks*), and *uncover the root causes of bottlenecks* for performance optimization.

Although several existing tools can automate analysis processes to some extent, previous work fails to resolve this issue in several ways. First, in the traditional approaches for performance debugging, though the processes of data collection are often automated, analysis and code optimization need great manual efforts. Application programmers need to learn the appropriate tools, which are often inundated with volumes of metrics, complexly presented graphs and tables [7], and apply their expertise to interpret data and its relation to the code [27] so as to optimize the code. For example, HPCViewer [14] and TAU [36] display the performance metrics through a graphical interface to users. Users apply their expertise to choose valuable data, which is hard and tedious for users, especially for non-expert users.

Second, *with apriori knowledge*, previous work proposes several automatic analysis solutions to identify critical bottlenecks. For example, in EXPERT [2] [3] [4] [5], *known performance problems* are specified in terms of execution patterns that represent situations of inefficient behavior. Analysis of trace data is done using an automated pattern-matching approach [5]. The Paradyn parallel performance tool [7] starts searching for bottlenecks by issuing instrumentation requests to collect data for a set of *pre-defined performance hypotheses* and comparing the collected performance data to *predefined thresholds*. Instances where the measured value for the hypothesis exceeds the threshold are defined as bottlenecks [9]. Using decision tree classification, which is trained by microbenchmarks that demonstrate both efficient and inefficient communication, the work [28] automatically classifies individual communication operations and reveals the cause of communication inefficiencies in the application.

Third, most of existing tools [16][22][23][24][27][28][29] take exploratory or confirmatory data analysis approaches to automated discoveries of relationships of relevant performance data. However, these efforts do not focus on locating performance bottleneck and uncovering their root causes of different performance bottlenecks. For example, PerfExplorer [35] uses clustering and dimensionality reduction techniques to manage large-scale data complexity, and performs comparative and correlation analysis techniques for automated discovery of relevant data relationships. The methodology of the work of [16] proposes a top down methodology towards automatic performance analysis of parallel applications, and uses clustering techniques to summarize and interpret performance information by identifying patterns or groups of code regions characterized by a similar behavior.

In this paper, we automate the whole process of debugging performance problems of parallel programs. Our method is unique in terms of two dimensions: first, *without any apriori knowledge*, we automatically locate bottlenecks and uncover their root causes for performance optimization, while EXPERT, Paradyn, and KOJAK *need apriori knowledge*; second, our method is lightweight in terms of the size of collected and analyzed performance data. For example, if a program is divided into $n$ code regions and the total number of processes is $m$, then we only need collect and analyze the performance data with the size of $125*n*m$ bytes, while PerfExplorer collects lots of data. Among these data, only 33% are used to locate the bottlenecks and the remains are used to find out root causes of these bottlenecks.

Our contributions are concluded as follows:
- For SPMD parallel applications, we present two effective clustering and searching algorithms to locate and refine bottlenecks.
- We propose to use the rough set approach to automatically uncover the root causes of bottlenecks.
- We design and implement the *AutoAnalyzer* system. We use two production applications to verify the effectiveness and correctness of our methods. According to the analysis results of *AutoAnalyzer*, we optimize two parallel programs with performance improvements by minimally 20% and maximally 170%.

This paper has seven sections. Section 2 formulates the problem. Section 3 summarizes the approaches. Section 4 introduces the *AutoAnalyzer* implementation. Section 5 evaluates the system. Section 6 outlines the related work. Finally, Section 7 draws a conclusion.

## 2. Terminology and Problem statement

A *code region* is a section of code that is executed from start to finish *with one entry and one exit*. For example, a code region can be *a function*, *subroutine* or *loop*. A code region can be nested within another one.

In our system, *a bottleneck is a code region that takes up a significant proportion of program's running time and has the potential for performance improvement*. If a code region takes up a trivial proportion of program's running time, the performance improvement of the code region will contribute little to the overall performance of the program, so we do not consider it as a possible bottleneck.

As shown in Figure 1, code regions are organized as a tree structure with the whole program as the root. According to the definition of the tree data structure, for any node $n_i$, the *depth* of $n_i$ is the length of the unique path from the root to $n_i$. For example, in Figure 1, the depth of *code region 1* is 1.

In our methodology, the code regions that have the same depth can not be overlapped. We encourage the nesting of code regions because deep nesting leads to fine granularity, which is helpful in narrowing the scope of the source code in locating bottlenecks. In the code region tree, for a node, their children nodes are nested within it. For example, in Figure 1, *code region 1* and *code region 2* do not intersect. *Code region 4* and *code region 6* are nested within *code region 1*.

We call the *code region that is a bottleneck* a *Critical Code Region (CCR)*. A CCR code region with the depth of $L$ is called an *L-CCR*.

When a CCR satisfies the following conditions, we call it *a Core of Critical Code Region (CCCR)*: (1) the CCR is a leaf node in the code region tree; (2) For a CCR, its children nodes are not CCR.

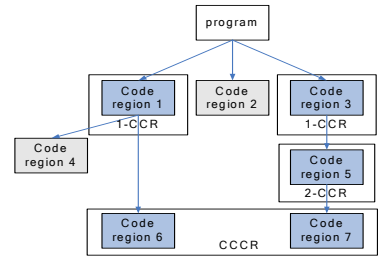

**Figure 1.** The code region tree of a parallel program.

We classify bottlenecks into two categories: *internal bottlenecks* and *external bottlenecks*. Specifically, internal bottlenecks occur in a local process or thread, caused by root causes such as poor data locality, poor efficiency in I/O operation and inefficient computing algorithm; external bottlenecks occur in interactions among different processes and threads, caused by the root causes such as load imbalance and resource contention.

Because of large quantity of performance data [7], which is produced by profiling or tracing methods, it is tedious and difficult to find out bottlenecks through manual efforts.

Our methods aim to solve the following problems in a fully automatic way. First, are there any bottlenecks in the program? Second, where are bottlenecks? Last, what are the root causes of bottlenecks?

## 3. Solutions

In this section, first, we give out the overview of our method, and then present two algorithms for locating external and internal bottlenecks respectively. Finally, we propose the rough set method to uncover their root causes.

### 3.1 Overview

Our method includes instrumentation, collection of performance data, and analysis of bottlenecks.

First, we instrument a whole parallel program into *code regions*. Our tool uses source-to-source transformation to automatically insert instrumentation code, and *requires almost no human involvement*.

Second, we need to collect performance data of parallel programs. For investigating the existence of bottlenecks and locating bottlenecks, we only collect the following performance data of each code region in different processes, which includes application layer performance data: *wall clock time* and *CPU clock time,* and hardware counter layer performance data: clock cycle and the quantity of instructions.

Besides, we collect the following performance data for uncovering the root causes for bottlenecks: parallel interface layer performance data includes *MPI communication time and quantity*; operation system layer performance data includes *disk I/O quantity;* hardware counter layer performance data includes *L1 cache miss, L2 cache miss, L1 cache access, L2 cache access*.

Based on the hardware counter layer performance data, we obtain two derived metrics, *L1 cache miss rate* and *L2 cache miss rate. For example* L1 cache miss rate can be obtained according to the formula *L1 cache miss / (L1 cache miss + L1 cache access)*.

Third, we investigate the existence of bottlenecks with two clustering algorithms. After we confirm the existence of bottlenecks, we use two different searching algorithms to search CCCRs, which are the locations of bottlenecks. After we have identified bottlenecks, we use the rough set method to analyze the detailed performance data of bottlenecks to find their root causes.

### 3.2 External Bottlenecks Detection

#### 3.2.1 Existence of external bottlenecks

For SPMD programs, each process or thread executes the same code. If we exclude code region in the master process responsible for the management routines, *the high similarity degree in behaviors of all processes or threads indicates the balance of workload dispatching and resources utilizing, and vice versa [16]*. So we use the similarity analysis method to investigate the possibility of existence of external bottlenecks. The performance similarity is analyzed among all participating computing processes or threads to discover the discrepancy.

We choose the *CPU clock time* of each code region as the main measurement. Different from the *wall clock time*, CPU clock time measures only the time during which the processor is actively working on a certain task, while the wall clock time measures the total time for a process to complete. Using CPU clock time, we exclude the time that passes due to communication delays or waiting for resources to become available.

We presume that the whole program is divided into *n* code regions, and the whole program has *m* processes or threads.

In our approach, each process or thread's performance is represented by a vector $\vec{V}_i$, where $i$ is the rank of process or thread, and the CPU clock time of *code region t*, $\vec{T}_{it}$, is the $t_{th}$ dimension of the vector. So $\vec{V}_i$ is described as $\vec{V}_i = <T_{i1}, T_{i2} \cdots, T_{in}>$.

We define the Euclidean distance $Dist_{ij}$ of two vectors $\vec{V}_i$ and $\vec{V}_j$ in Equation (1).

$$Dist_{ij} = \sqrt{(T_{i1} - T_{j1})^2 + \cdots + (T_{in} - T_{jn})^2} \quad (1)$$

Based on the calculation of *$Dist_{ij}$*, an OPTICS clustering method [1] is used to classify all processes or threads. We choose the OPTICS clustering method because it has advantage in discovering isolated points. In this approach, the performance vector of each process or thread is considered as a point in *an n-dimension space*. And a set of points are classified into one *cluster* when the point density in the area, where these point scattered, is larger than a threshold. Figure 2 presents the OPTICS clustering algorithm. If a point is not included into any cluster, then we consider it as an isolated point. Each isolated point is a cluster.

---

```
count=0; select a performance vector V_p;
for each point V_q in the n-dimension space
{
  if (distance(V_p , V_q)<threshold) then count++;
  // we set the threshold as 10% * len (V_p)
}
if (count>count_threshold) //count_threshold=2
    confirm there is a cluster and select another point
    not in the cluster
else
    select another point
repeat the top line until all vectors are compared
```

---

Figure 2. The OPTICS clustering algorithm.

For SPMD programs, if all processes have similar performance behaviors and only one cluster is finally obtained, we confirm that there is no external bottleneck. Otherwise we presume that external bottlenecks exist in the program.

We propose a *metrics of S* to measure the severity degree of dissimilarity of a program, which is calculated according to Equation (2) and Equation (3). The larger S means more severe degree in performance dissimilarity among all processes or threads.

$$S = \frac{\max(Dist_{ij})}{\min(len_i)} \quad (2)$$

$$len_i = \sqrt{T_{i1}^2 + ... + T_{in}^2} \quad (3)$$

#### 3.2.2 Locating external bottlenecks

When users confirm the existence of external bottleneck, they need to further locate external bottlenecks. In this section, we propose a top-down algorithm to search external bottlenecks.

We presume there are *r* code regions whose depth is *one* in code region tree of *n* code regions.

The algorithm of searching external bottlenecks is as follows:

**Step 1**. For each *code region c*, *c=1…n*, if its depth is great than 1, we set $T_{ic}$ =0 in each performance vector $V_i$, $i=1…m$.

**Step 2**. For each *code region k* whose depth is *one*, if $T_{ik}$ >0 $(i=1…m)$, we set $T_{ik}$ as 0 and watch the change of the output of the OPTICS clustering one by one. If the output of the algorithm changed, for example, *the number of clusters* or *members of a cluster* changed, we confirm *code region k* is a 1-CCR and include it into the set of 1-CCRs, because it can influence the clustering result. After having obtained all 1-CCRs, we restore the original value for each $T_{ic}$, where $i=1…m$ and $c=1…n$, and then go to Step 3. If no 1-CCR is obtained, we go to Step 5.

**Step 3**. For each *code region p* that is a *(L-1)-CCR* ($L \geq 2$), and its each child node *k,* we set $T_{ip}$ and $T_{ik}$ as 0 ($i=1…m$). For each *code region k*, we restore $T_{ik}$ with the original CPU clock time in each vector $T_i$, and use the OPTICS clustering algorithm to classify the vectors. If the output of the algorithm does not change, we confirm that *code region k* is an *L*-CCR.

**Step 4**. We recursively analyze the children of *code region k*. Repeat Step 4 until the CCR *k* is a leaf node or any one of its children is not CCR. Then *code region k* will be added to the set of CCCR. The algorithm ends.

**Step 5**. We combine *s* code regions of the depth of *one*, where *s=2*, into one composite code region. The number of composite code region is $C_R^s$. We calculate the new performance vectors and use the clustering algorithm to obtain the new clustering result, which is the new reference for searching CCCRs. We create the new code region tree with the composite code regions as the nodes of the depth of *one*, and repeat Step 1 to Step 4 to search the set of CCCR.

When *s=2*, if we do not find out any 1-CCR, then we increment (s<=r-1), and continue Step 5 until s=r-1.

According to Step 3 in our searching algorithm, the CCCR has dominated effect on the clustering result in comparison with its parent CCR. So for SPMD programs, we only consider that CCCRs are the set of external bottlenecks, which users should focus on to improve the performance.

### 3.3 Internal bottlenecks detections

#### 3.3.1 Normalize metrics

During the analysis, there is large quantity of performance data collected in each process. However, it is usually confused for users, especially non-expert ones, to decide which metrics are the most important. In this Section, we propose a single normalized metric, named *Code Region Normalized Metric (CRNM)*, as the measurement basis for locating internal bottlenecks and further measuring their severity.

For each code region, CRNM is defined in Equation (4):

$$CRNM = \frac{CRWT}{WPWT} * CPI \quad (4)$$

In Equation (4), *CRWT* is the wall clock time of a code region; *WPWT* is the wall clock time of the whole program; CPI is the average cycles per instruction of each code region.

We choose CRNM because of the following two reasons: first, by using *the ratio* of *the wall clock time of a code region* to *the wall clock time of the whole program*, users can judge the performance contribution of a code region to the overall performance of a program. Secondly, average cycles per instruction (*CPI*) measures the efficiency of instruction execution. CPI, which can be derived from the total instruction number and the total executing cycles, is a basic metric that can be affected by nearly all hardware events: cache or TLB miss, cache line invention, pipeline stall caused by data dependency or branches misprediction and so on. So our normalized CPI represents a measure of *the importance of a code region to the overall performance of the application.*

#### 3.3.2 Locating internal bottlenecks

The procedure of searching internal bottlenecks includes three steps:

First, for each processes or thread, we obtain the CRNM value of each code region. If a code region is not on the call path in a process, its CRNM value is zero. And then we obtain the average value of each code region for all processes or threads, taking into account that *SPMD programs can contain 'if' statements*.

Second, we use the k-means clustering method [12] to classify each code region according to the average CRNM value. This is because the k-means clustering method can classify the data into *k* clusters without the threshold value provided by users. We define five *severity categories*: *very high (4)*, *high (3)*, *medium (2)*, *low (1)*, and *very low (0)*. The k-means clustering method finally classifies each code region into one of severity categories according to its CRNM value. Figure 3 show the procedure of using k-means clustering method [12].

Third, if a code region is classified into one of the severity categories of *very high* or *high*, we consider it as a critical code region (CCR). .

In order to refine the scope of internal bottlenecks, we also propose a recursive searching algorithm in the code region tree, as follows:

(1) If a leaf node *j* is a CCR, then *code region j* is a core of critical code region (CCCR)

(2) For a none-leaf node *j*, if the severity degree of each child is less than its severity degree, then we consider *code region j* as a CCCR.

The set of CCCR is also the set of internal bottleneck. The severity degree of internal bottlenecks can be measured by the severity category of a code region.

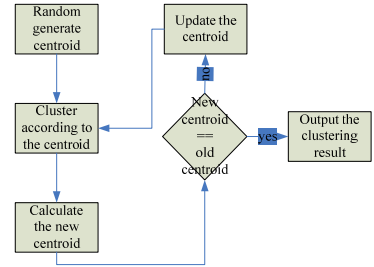

Figure 3. The k-means clustering procedure [12]. Step 1: partition data into k subsets with k random seeds as the initial centroids. Step 2: compute the centroids of each cluster of the current partition. Step 3: go back to Step 2, stop when no more new assignment.

### 3.4 Root cause Analysis

#### 3.4.1 The rough set approach [15] [19]

The rough set approach is a data mining method that can be used for categorizing, data association and so on. In this paper, we use the rough set approach to uncover the root causes of different bottlenecks.

We start with introducing some basic terms in the approach, including *decision table*, *core*, and then present one solution to find core.

As shown in Table 1, the *decision table* is used to describe the large amount of data. Each entry of a decision table consists of three parts: *entry ID*, *attributions*, and *decision*. For example, in Table 1, there are four attributions: *a1, a2, a3, and a4*.

*Core* is a special set of attributions that is critical in distinguishing the decisions. How to find the core and remove the trivial attributions in determining the decision is a main research field in the rough set approach. One of the solutions is to create *discernibility matrix* [15] according to the decision table, and extract the core using discernibility matrix.

The process of creating discernibility matrix is as follows:

Assume the set of entry IDs in the decision table is $\{x_1, x_2, \cdots, x_n\}$.

*A* is the set of attributions and *D* is the set of decisions. For entry $x_i$, we define a ($x_i$) as the value of attribution *a*, $a \in A$. According to Equation (5), we calculate the value of each element $c_{ij}$ in the discernibility matrix.

$$c_{ij} = \begin{cases} \{a \in A : a(x_i) \neq a(x_j)\} & D(x_i) \neq D(x_j) \\ 0 & D(x_i) = D(x_j) \quad i,j = 1\cdots n \\ -1 & a(x_i) = a(x_j) \quad D(x_i) \neq D(x_j) \end{cases} \quad (5)$$

For Table 1, the calculated discernibility matrix is shown in Figure 4. Because every discernibility matrix is symmetrical, we only consider its upper triangular part.

Table 1. An example of decision table.

| ID | a1 | a2 | a3 | a4 | decision |
|---|---|---|---|---|---|
| 0 | sunny | hot | high | False | N |
| 1 | sunny | hot | high | True | N |
| 2 | overcast | hot | high | False | P |
| 3 | sunny | cool | low | False | P |

$$\begin{pmatrix} 0 & 0 & a1 & a2a3 \\ & 0 & a1a4 & a2a3a4 \\ & & 0 & 0 \\ & & & 0 \end{pmatrix}$$

Figure 4. The discernibility matrix for the decision table in Table 1.

Using discernibility matrix, extracting the core attributions is much easier and the algorithm is as follows:

**Step 1.** We exclude the elements in the discernibility matrix whose value is 0 or -1 and consider the ones with the value of a set of attributions. If the element owns the value of attribution set which contains only one attribution, we add this attribution into the core set (CS), because it is a core attribution that plays a critical role in making decisions. For example, the CS for Figure 4 is *{a1}*.

**Step 2.** If the value of element in the discernibility matrix, for example *{a2a3}* in Figure 4, does not contain any attributions in the CS, we change the CS as a conjunctive normal form, CS ∧ {a2a3}. In the example, the CS is finally as *{a1} ∧ {a2a3}*.

**Step 3.** We transform the CS from a conjunctive normal form into a disjunctive normal form. Then we select the conjunctive minor which owns the least number of attributions and occurs in the most times. These attributions are the finally core of the rough set approach. In our example, the core is *{a1, a2}* or *{a1, a3}*.

### 3.4.2 Root causes of external bottlenecks

For performance optimization, users need to know the root causes of bottlenecks. In this section, we propose the rough set method to uncover the root causes of external bottlenecks, and give hints for performance improvements. In order to decrease the size of the performance data, the collected and analyzed attributions are confined to the performance metrics of CCCRs.

As shown in Figure 5, we create the decision table as follows:

We choose the rank of each process as the entry ID. We select *L1 cache miss rate*, *L2 cache miss rate*, *disk I/O quantity*, *network I/O quantity* and *executing instruction number* as five different attributions a(i), i=1..5.

For *process k*, where *k* is the ID, the element of the decision table corresponding to *a (i)* is obtained as follows:

We take attribution *a (1)* (L1 cache miss rate) as an example:
For performance vector $T_k$, where k=1...m, $T_{kt}$ is assigned with the value of *L1 cache miss rate* of code region *t* in process *k*.

After having created the performance vector, we use the OPTICS clustering algorithm to obtain clusters. If $T_k$ is classified into *the cluster with the ID of 0*, we assign this ID to the element of the decision table corresponding to *a (1)*. For example, in Table 2, for *process 0* and *process 1*, the corresponding values of attribution *a (1)* are 0. That is to say, performance vector $T_0$ and performance vector $T_1$ are classified into the cluster with the ID of 0.

For *process k*, the decision value is the ID of the cluster into which the process *k* is classified according to the metrics of the CPU clock time.

How to extract the core from the attributions by using discernibility matrix method is introduced in Section 3.4.1. The core set is the major root causes that cause the dissimilarity of behaviors of processes.

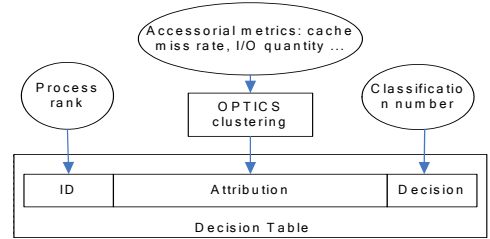

Figure 5. The approach of uncovering the root causes of external bottlenecks.

### 3.4.3 Root Causes of Internal Bottlenecks

In this section, we also use the rough set method to explore the root causes of internal bottlenecks. We create the decision table as follows:

We use code region ID as *the Entry ID* to identify each table entry; we also select *L1 cache miss rate*, *L2 cache miss rate*, *disk I/O quantity*, *network I/O quantity* and *executing instruction number* as five different attributions.

For *code region k*, the element of the decision table corresponding to *a (i)* is obtained as follows:

We take attribution *a (1)* (L1 cache miss rate) as an example:

We obtain the average value of L1 cache miss rates of each code region in different processes or threads. We use the K-means clustering algorithm to classify the average L1 cache miss rate of each code region into five categories: very high (4), high (3), medium (2), low (1), and very low (0). For *code region k*, if its severity category is higher than medium, we assign 1 to element corresponding to the attribution a (1), otherwise 0.

Last, for *code region k*, if it's an internal bottleneck according to the approach proposed in Section 3.3.2, then the decision value is 1, otherwise it is 0.

After creating the decision table, we extract core sets using the discernibility matrix, described in Section 3.4.1. Because the core sets are the attributions that affect the decision, we confirm that attributions in core sets are the root causes of internal bottlenecks.

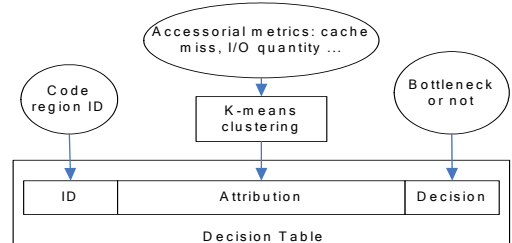

Figure 6. The approach of uncovering the root causes of internal bottlenecks.

## 4. AutoAnalyzer Implementation

In order to evaluate the effectiveness of our proposed methods, we have designed and implemented a prototype, *AutoAnalyzer*. Up to now, *AutoAnalyzer* supports the debugging of performance problems of SPMD style MPI applications, written in C, FORTRAN 77 and FORTRAN 90. Figure 7 shows the architecture of AutoAnalyzer.

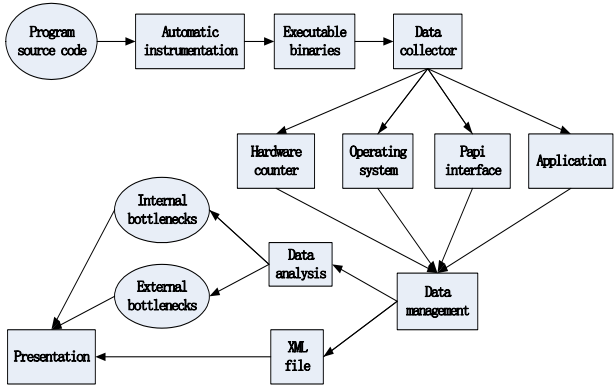

Figure 7. The architecture of AutoAnalyzer.

The major components of *AutoAnalyzer* include *automatic instrumentation*, *data collector*, *data management*, and *data analysis*.

*Automatic instrumentation.* On the basis of OMPi [17], a source-to-source compiler, we have implemented source code-level instrumentation. Our tool uses source-to-source transformation to automatically insert instrumentation code and *almost requires no human involvement*. After parsing the program, abstract syntax tree (AST) is built. AST shows program's structure information. For example, the beginning and ending position of *functions, procedures or loops*. With this information, our instrumentation codes can be automatically inserted to define code regions.

Furthermore, users can conveniently interact with the tool through its interface to instrument the code by eliminating, fusing and splitting code regions because users can cope with source codes to optimize the instrumentation granularity. Users can determine where to insert instrumentation codes by changing the parameters of the interface，which determines several modes: *outer loop*, *inner loop*, *mathematical library*, *parallel interface library*, for example MPI, system call, C / FORTRAN library and user defined functions or procedures (for FORTRAN). Users can choose one or more modes to instrument the code.

Without any restriction in the instrumentation, the program can be divided into hundreds or thousands of code regions. For example, after instrumentation, a parallel program with only 2,000 lines is divided into more than 300 code regions. This situation has negative influence in the performance analysis because large amount of data are collected. Moreover large number of instrumentation codes disturb the program's performance. Thus, in order to reduce the number of code regions, we only select the modes of user defined functions or procedures, and outer loops to instrument, so we can decrease the number of code regions remarkably.

*Data collector.* We collect performance data from four hierarchies: application, parallel interface, operating system and hardware.
In the application hierarchy, we collect wall clock time and CPU clock time of each code region. In the parallel interface hierarchy, we have implemented an MPI library wrapper to record MPI routines' behaviors of both point-to-point and collective communication. The wrapper is implemented by wrapping the interface of MPI library, PMPI, instead of MPI's routines. In the wrapper, we instrumented codes to collect information of MPI library, for example, the quantity of messages and the time consumed in MPI library.

In the operating system hierarchy, we use systemtap [20] to monitor disk I/O and record the quantity and time of I/O operations. Systemtap is based on Kprobe, which is implemented in Linux kernels. Kprobe can instrument system calls of Linux kernel to obtain the consuming time and functions' parameters as well as I/O quantity.

In the hardware hierarchy, we use PAPI [18] to count hardware events, including L1 cache miss, L1 cache access, L2 cache miss, L2 cache access, TLB miss and the number of executed instructions.

*Data management.* We collect all performance data instrumented in different nodes onto one node for analysis. All data are stored in an XML file

*Data analysis.* We use our innovative methods, such as k-means, OPTICS clustering algorithm, top-down searching method and the rough set approach to analyze all the performance data to search both internal bottlenecks and external bottlenecks.

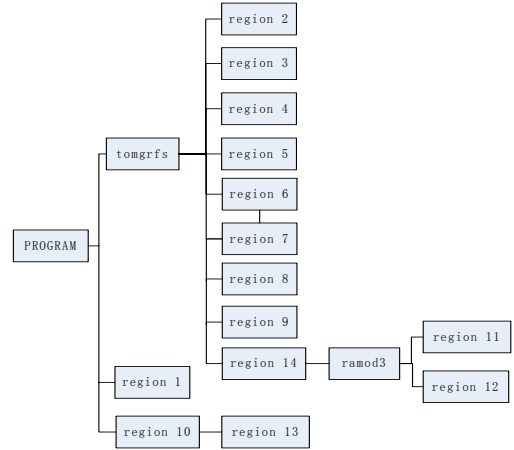

Figure 8. Code region tree of ST, a production application. Code region 11, 12 are in subroutine ramod3, which is nested in code region 14. All code regions contain loops.

## 5. Evaluation

In this section, we use two production parallel applications to evaluate the correctness and effectiveness of *AutoAnalyzer*. One program is *ST*, which calculates seismic tomography using the refutations method and *is on the production use in the largest oil company in China*. The other is the paralleled NPAR1WAY module of SAS, written in FORTRAN, which is a system widely used in data and statistical analysis.

### 5.1 ST

In this Section, we use a production parallel application of 4307 line codes, *ST*, to evaluate the effectiveness of our system. To identify a problem, a user of our tools does little to start. The tool automatically instruments the code. After analysis, the tool informs the user about bottlenecks and their root causes. For *ST*, it took about 2 days for an MS student in our lab to locate bottlenecks and rewrite about 200 lines to optimized codes.
In the rest of this section, we give the detail of locating bottleneck and optimizing performance.

Without any restriction on automatic instrumentation, *ST* is instrumented into about 300 code regions. This situation has negative influence on performance analysis because large amount of data are collected, and large number of instrumentation codes disturb the program's performance. To reduce the number of code regions, we support another way of instrumentation that allows a user to select whether to instrument *functions* or *procedures*, or *outer loops*. With this way, we instrument *ST* into 14 code regions, and Figure 8 shows the code region tree.

Our platform is 4 SMP machines with 8 CPUs, connected with Fast Ethernet 1000Mbps. The CPUs are AMD Opteron with 64KB L1 data cache, 64KB L1 instruction cache and 1MB L2 cache. The OS version is linux-2.6.19.

---

Performance similarity
there are 5 kinds of processes
kind 0: 0
kind 1: 1  2
kind 2: 3
kind 3: 4  6
kind 4: 5  7
dissimilarity severity, S: 0.783958
CCCR: region 11
CCR tree:
region 14 (1-CCR)  ---> region 11 (2-CCR & CCCR)

---

Figure 9. The analysis result of similarity measurement.

### 5.1.1 External bottleneck analysis

According to the approach proposed in Section 3.2, the similarity analysis result is displayed in Figure 9. We can find that all processes are classified into 5 clusters and the severity degree S is 0.783958. For SPMD programs, the analysis result indicates that external bottlenecks exist. According to the searching result, we can conclude that *code region 11* and *code region 14* are CCRs and *code region 11* is a CCCR, which is the location of the problem.

We create the decision table to analyze the root causes of *code region 11*.

Table 2. Decision table for the external bottlenecks.

| ID | a1 | a2 | a3 | a4 | a5 | D |
|----|----|----|----|----|----|---|
| 0  | 0  | 0  | 0  | 0  | 0  | 0 |
| 1  | 0  | 0  | 0  | 0  | 1  | 1 |
| 2  | 0  | 0  | 0  | 0  | 1  | 1 |
| 3  | 1  | 0  | 0  | 0  | 2  | 2 |
| 4  | 0  | 1  | 0  | 0  | 3  | 3 |
| 5  | 1  | 1  | 0  | 1  | 4  | 4 |
| 6  | 1  | 2  | 0  | 1  | 3  | 3 |
| 7  | 1  | 2  | 0  | 0  | 4  | 4 |

Table 2 shows the decision table. In the decision table, the attributions $a(i)$ ($i=1,2,3,4,5$) respectively represents L1 cache miss rate, L2 cache miss rate, disk I/O quantity, network I/O quantity and executing instruction number. Figure 10 shows the discernibility matrix.

$$\begin{bmatrix} 0 & a5 & a5 & a1a5 & a2a5 & a1a2a4a5 & a1a2a4a5 & a1a2a5 \\ & 0 & 0 & a1a5 & a2a5 & a2a4a5 & a1a2a4a5 & a1a2a5 \\ & & 0 & a1a5 & a2a5 & a2a4a5 & a1a2a4a5 & a1a2a5 \\ & & & 0 & a1a2a5 & a2a4a5 & a2a4a5 & a2a5 \\ & & & & 0 & a1a4a5 & 0 & a1a2a5 \\ & & & & & 0 & a2a5 & 0 \\ & & & & & & 0 & a4a5 \\ & & & & & & & 0 \end{bmatrix}$$

Figure 10.  Discernibility matrix for external bottleneck.

By using extracting algorithm in the discernibility matrix, we can get the core set {a5}, which indicates that the executing instruction of *code region 11* is the root cause for the dissimilar behavior of processes.

Figure 11 verifies our analysis from which we can discover the obvious difference in executing instruction quantities of code region 11 among different processes.

### 5.1.2 Internal bottleneck analysis

For *ST*, AutoAnalyzer outputs the CRNM of each code region in each process, and then obtain the average CRNM of each code region as shown in Figure 12. Using K-means clustering, we can obtain the result of k-means clustering method.

As shown in Figure 13, the severity degree of region 14, 11, 8 is larger than *medium*. According to the analysis result, we confirm code regions 14, 11 and 8 are critical code regions.

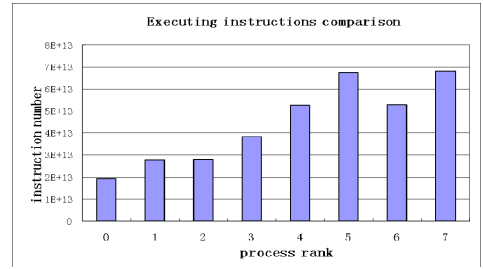

Figure 11. Variance of executing instructions of *code region 11* in different processes.

According to the algorithm in Section 3.3.2, since code region 11 is nested in *code region 14* and the severity degree of *code region 11* is the same as *code region 14*, so *code region 11* is a CCCR; no code region is nested in *code region 8*, so *code region 8* is also CCCR. We confirm that *code region 8* and code region 11 are internal bottlenecks.

We analyze the root causes of internal bottlenecks with the rough set approach. The decision table is created in Table 3. In the decision table, the attributions $a(i)$ ($i=1,2,3,4,5$) respectively represents L1 Cache miss rate, L2 cache miss rate, disk I/O quantity, network I/O quantity and executing instruction number. The obtained discernibility matrix is shown in Figure 14.

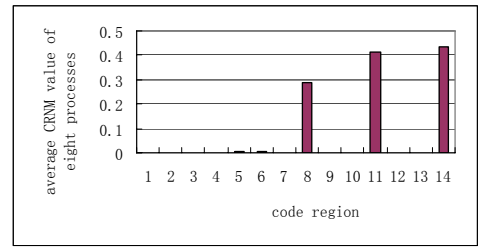

Figure 12. The CRNM value of 14 code regions.

```
very high: 14, 11
high: 8
medium: 5, 6
low: 2
very low: 1, 9, 3, 7, 10, 12, 13, 4
```

Figure 13. The result of k-means clustering. All code regions are classified into different 5 clusters.

Table 3. Decision table in the analysis of internal bottlenecks. We can see the core set {a2, a3}.

| ID | a1 | a2 | a3 | a4 | a5 | D |
|----|----|----|----|----|----|----|
| 1  | 0  | 0  | 0  | 0  | 0  | 0 |
| 2  | 1  | 0  | 0  | 0  | 0  | 0 |
| 3  | 0  | 0  | 0  | 0  | 0  | 0 |
| 4  | 0  | 0  | 0  | 0  | 0  | 0 |
| 5  | 1  | 1  | 0  | 0  | 1  | 0 |
| 6  | 1  | 0  | 0  | 0  | 1  | 0 |
| 7  | 0  | 0  | 0  | 0  | 0  | 0 |
| 8  | 0  | 0  | 1  | 0  | 1  | 1 |
| 9  | 1  | 0  | 0  | 0  | 0  | 0 |
| 10 | 1  | 0  | 0  | 0  | 0  | 0 |
| 11 | 1  | 1  | 0  | 0  | 1  | 1 |
| 12 | 0  | 0  | 0  | 0  | 0  | 0 |
| 13 | 0  | 0  | 0  | 0  | 0  | 0 |
| 14 | 1  | 1  | 0  | 0  | 1  | 1 |

By using extracting algorithm in Section 3.4.1, we can get the core set {a2, a3}. This means that L2 cache miss and disk I/O quantity are the root causes of internal bottlenecks. Then we search in the decision table and find that the root cause of *code region 8* is high disk I/O quantity and the root cause of *code region 11* is high L2 cache miss problem, which can be confirmed by the performance data. From the performance data, we can observer that the disk I/O quantity in *code region 8* is as high as 106G and the L2 cache miss rate in *code region 11* is 17.8%.

$$\begin{bmatrix}
0 & 0 & 0 & 0 & 0 & 0 & 0 & a_3a_5 & 0 & 0 & a_1a_2a_5 & 0 & 0 & a_1a_2a_5 \\
 & 0 & 0 & 0 & 0 & 0 & 0 & a_3a_5 & 0 & 0 & a_2a_5 & 0 & 0 & a_2a_5 \\
 & & 0 & 0 & 0 & 0 & 0 & a_1a_2a_5 & 0 & 0 & a_1a_2a_5 & 0 & 0 & a_1a_2a_5 \\
 & & & 0 & 0 & 0 & 0 & a_3a_5 & 0 & 0 & a_1a_2a_5 & 0 & 0 & a_1a_2a_5 \\
 & & & & 0 & 0 & 0 & a_3 & 0 & 0 & 0 & 0 & 0 & 0 \\
 & & & & & 0 & 0 & a_3 & 0 & 0 & a_2 & 0 & 0 & a_2 \\
 & & & & & & 0 & a_3a_5 & 0 & 0 & a_1a_2a_5 & 0 & 0 & a_1a_2a_5 \\
 & & & & & & & 0 & a_1a_3a_5 & a_1a_3a_5 & a_1a_2 & a_3a_5 & a_3a_5 & 0 \\
 & & & & & & & & 0 & 0 & a_2a_5 & 0 & 0 & a_2a_5 \\
 & & & & & & & & & 0 & a_2a_5 & 0 & 0 & a_2a_5 \\
 & & & & & & & & & & 0 & a_1a_2a_5 & a_1a_2a_5 & 0 \\
 & & & & & & & & & & & 0 & 0 & a_1a_2a_5 \\
 & & & & & & & & & & & & 0 & a_1a_2a_5 \\
 & & & & & & & & & & & & & 0
\end{bmatrix}$$

Figure 14. Discernibility matrix for decision table in Table 3.

### 5.1.3 The performance optimization

In Section 5.1.1, we confirm that *code region 11* is external bottleneck, caused by variance of executing instructions of *code region 11* in each process.

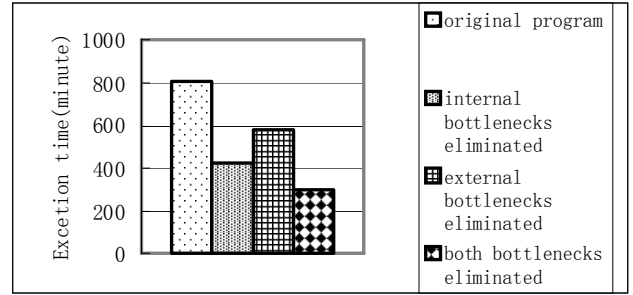

Figure 15. Performance of program before and after optimization.

In order to eliminate the external bottlenecks, we tune the workload in *code region 11* by dynamic dispatching using a master process, instead of taking a static way as the original. After the optimization, we use *AutoAnalyzer* to analyze the optimized code again. All computing processes are classified into one cluster, and the *severity degree of dissimilarity* S decreases from 0.783958 to 0.032800, which means that all processes have the similar performance with balanced work loads.

In Section 5.1.2, we confirm that *code region 8* and *code region 11* are internal bottlenecks, which is respectively caused by high disk I/O quantity and high L2 cache miss.

We take the following approaches to optimize the code. First, we improve *code region 8* by buffering as many as data into the memory. Second, we improve the data locality of *code region 11* by breaking the loops into small one and rearranging the data storage.

We use *AutoAnalyzer* to analyze the optimized code again. *Code region 8* is not the bottleneck any longer, while *code region 11* is still the internal bottlenecks. However, the average CRNM value of *code region 11* decreases from 0.41 to 0.26. The root cause is no longer the high L2 caches miss rate, but the large quantity of executing instructions.

Figure 15 shows the performance of *ST* before and after optimization. With internal bottleneck eliminated, the performance of the optimized *ST* rises by 90% in comparison with the original program. With external bottleneck eliminated, the performance of the optimized *ST* rises by 40% in comparison with the original program. With both internal and external bottlenecks eliminated, the performance of the optimized program rise by 170% in comparison with the original program.

### 5.2 NPAR1WAY

We instrument the whole program of NPAR1WAY into 12 code regions to separate *functions, subroutines and outer loops*. Our platform is a cluster system, and each CPU is Intel® Xeon® Processor E5335, which is quad core with 2 GHz CPU speed, 8 MB L2 cache size, 2 GHz L2 cache speed. The operating system is Linux 2.6.19.

#### 5.2.1 External bottleneck analysis

According to the output of AutoAnalyzer, the similarity analysis result is displayed in Figure 16. We can find that all processes are classified into one cluster, which indicates that no external bottleneck exists.

```
Performance similarity
there is 1 kind of processes
kind 0: 0  1  2  3  4  5  6  7
```

Figure 16. The analysis result of similarity measurement.

### 5.2.2 Internal bottleneck analysis

*AutoAnalyzer* outputs the CRNM of each code region in each process, and then obtains the average CRNM of each code region as shown in Figure 17. Using the K-means clustering approach, we obtain the analysis result. The clustering result in Figure 18 shows that the severity degree of *code region 3*, and *code region 12* are larger than *medium*. So *code region 3* and *code region 12* are CCRs. Because there is no nested code regions in *code region 3* and *code region 12*, both of the two code regions are CCCRs, which we consider internal bottlenecks.

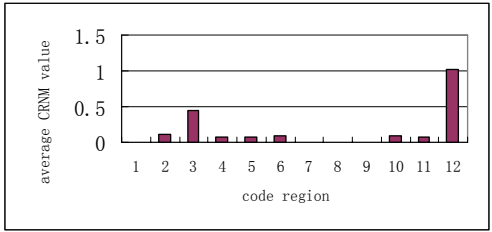

Figure 17. The average CRNM value of code regions of eight processes.

The root causes of internal bottlenecks are analyzed by the rough set approach and the decision table is created. In the decision table, the attributes a(i) (i=1,2,3,4,5) respectively represents *L1 cache miss rate, L2 cache miss rate, disk I/O quantity, network I/O quantity and executing instruction number*.

By using the extracting algorithm in the discernibility matrix, we can get the core set {a4, a5}. This indicates that both *high communication data quantity* and *high instruction number* are root causes of the internal bottlenecks. Then we search in the decision table and find that *code region 3* has high executing instruction number, at the same time *code region 12* has high executing instruction number and high network I/O quantity. From the performance data, we can see that executing instruction of *code region 3* and *code region 12* are 26% and 60% of total executing instruction of the program respectively and network I/O quantity of *code region 12* is 70% of the total network I/O quantity.

```
very high: 12
high: 3
medium: 6, 10, 2
low: 4, 11, 5
very low: 1, 7, 8, 9
```

Figure 18. The result of k-means clustering. All code regions are classified into different 5 clusters.

### 5.2.3 The performance optimization

According to the root causes uncovered by *AutoAnalyzer*, we adopt several approaches to optimize the codes to eliminate internal bottlenecks.

We improve the performance of *code region 3* and *code region 12* to relieve the severity of internal bottlenecks. We optimize *code region 3* and *code region 12* by eliminating *redundant common expressions*. For example, there is one common multiply expression occurring three times in *code region 3*. We use one variable to store the result of the multiply expression at its first appearance and directly use the variable to avoid redundant computation subsequently. In this way, we can decrease massive instructions by eliminating redundant common expressions in deep loops.

Then we re-analyze the codes. For the optimized *code region 3*, the analysis results show that the total number of executing instructions and the wall clock time are reduced by 36.32% and 20.33% respectively. For the optimized *code region 12*, the analysis results show that the total number of executing instructions and the wall clock time are reduced by 16.93% and 8.46% respectively. For *code region 12*, we fail to eliminate high network I/O quantity.

Figure 19 shows the performance improvement. The performance of the paralleled NPAR1WAY rises by 20% after optimization.

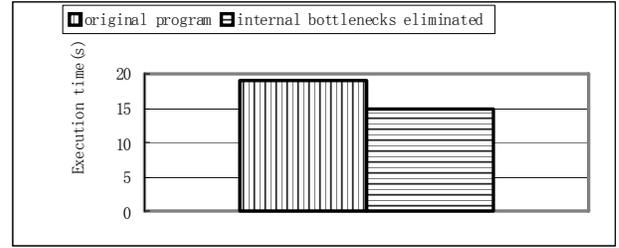

Figure 19. Performance improvement of the program before and after optimization.

## 6. Related work

The traditional approach for performance debugging is through the processes of data collection, which is often automated, analysis and code optimization that need great manual efforts.

In this approach, application programmers need to learn the appropriate tools, which are inundated with volumes of metric, complexly presented graphs and tables [7], and rely their expertise to interpret data and its relation to the code [27] so as to optimize the code. For example, HPCViewer [14] and TAU [36] display the performance metrics through a graphical interface to users. Users depends their expertise to choose valuable data, which is hard and tedious for users, especially non-expert users.

With *apriori knowledge*, previous work proposes many automatic analysis solutions to identify critical bottlenecks. EXPERT [2] [3] [4] [5] describes performance problems using a high level of abstraction in terms of execution patterns that result from an inefficient use of the underlying programming model(s). *Known performance problems* are specified in terms of execution patterns that represent situations of inefficient behavior. Analysis of trace data is done using an automated pattern-matching approach [5]. The Paradyn parallel performance tool [7] starts searching for bottlenecks by issuing instrumentation requests to collect data for a set of *pre-defined performance hypotheses* for the whole program. Paradyn starts its search by comparing the collected performance data to *predefined thresholds*. Instances where the measured value for the hypothesis exceeds the threshold are defined as bottlenecks [9]. Paradyn starts a hierarchical search of the bottlenecks and refines this search by using stack sampling [11] and by pruning the search space considering the behavior of the application during previous runs [8]. Using decision tree classification, which is trained by microbenchmarks that demonstrate both efficient and inefficient communication, the work [28] automatically classifies individual communication operations and it reveals the cause of communication inefficiencies in the application. Aksum [34] automatically performs multiple runs of a parallel application and detects performance bottlenecks by comparing the performance achieved varying the problem size and the number of allocated processors. The key idea in the work of [38] [39] is to extract performance knowledge from parallel design patterns or model that represent structural and communication patterns of a program for performance diagnose. The distinguished difference of our *AutoAnalyer* is that we automatically locate bottlenecks and uncover their root causes for performance optimization *without apriori knowledge*.

Many previous efforts propose automatic performance analysis approaches, for example exploratory or confirmatory data analysis [27] or fuzzy set method [31] to automated discoveries of relevant performance data.

PerfExplorer [22][23][24] addresses the need to manage large-scale data complexity using techniques such as clustering and dimensionality reduction, and the need to perform automated discovery of relevant data relationships using comparative and correlation analysis techniques. By clustering thread performance for different metrics, PerfExplorer should discover these relationships and which metrics best distinguish their differences. In the work of [24], PerfExplorer inference rules have been developed to recognize and diagnose performance characteristics important for optimization strategies and modeling. The methodology of the work of [16] proposes a top down methodology towards automatic performance analysis of parallel applications, which first focuses on the overall behavior of the application in terms of its activities. Then, the individual code regions of the application and the activities performed within each of them are considered. Clustering techniques help in summarizing and interpreting this information by identifying patterns or groups of code regions characterized by a similar behavior. The work by Ahn and Vetter [29] chooses to use several multivariate statistical analysis techniques to analyze parallel performance behavior. The types of analysis they performed included cluster analysis and F-ratio, factor analysis, and principal component analysis. They showed how hardware counters could be used to analyze the performance of multiprocessor parallel machines. The primary goal of the work of SimPoint [32] is to reduce long-running applications down to tractable simulations. The authors define the concept of "basic block vectors", and use those concepts to define the behavior of blocks of execution, usually one million instructions at a time. The work of [30] [31] describes a fuzzy based bottleneck search, a performance similarity measure for code regions and experiment factors, and performance similarity analysis. However, it does not want to uncover the root causes of bottlenecks. The work of [17] proposes the approaches to measure and attribute parallel idleness and parallel overhead of multithreaded parallel applications.

The work of [35] describes a scalable and general-purpose framework for auto-tuning compiler-generated code, which combines Active Harmony's parallel search backend with the CHiLL compiler transformation framework to generate in parallel a set of alternative implementations of computation kernels and automatically select the one with the best-performing implementation.

Our *AutoAnalyzer* is similar to the above work in proposing performance vectors to represent behavior of parallel application [16][22][23][24] [30]and applying clustering algorithms to the investigation of existence of bottlenecks[16][22][23][24][27][29]. However, there are two distinguished different points: first, in addition to the automatic performance analysis, we also automatically locate bottlenecks and uncover their roots causes for performance optimization; second, our method is lightweight in terms of the collected and analyzed performance data. We propose simple and effective performance metrics to represent performance behavior of parallel applications, with which we clusters the performance vector and search bottlenecks in terms of code region. Only on condition that we want to uncover the root causes of bottlenecks, we need collect and analyze diversity of performance data of the possible bottlenecks.

The work [33] proposes a plan to develop a test suite for verifying the effectiveness of different tools in debugging performance problems. If it succeeds, this tool can provide a benchmark for evaluating the performance and precision of different tools in terms of the false positive and the false negative. Unfortunately, this project seems ended without updating its web site.

7. Conclusions

This paper presented a series of innovative methods in performance debugging of SPMD styled parallel programs in a fully automatic way. Our contribution is threefold: first, we proposed simple performance metrics to represent behavior of different processes of parallel programs, and presented two effective clustering and searching algorithms to locate both internal and external bottlenecks; second, we proposed the rough set algorithm to automatically uncover the root causes of bottlenecks; third, we designed and implemented the *AutoAnalyzer* system, and used two production applications to verify the effectiveness and correctness of our methods. According to the analysis results of *AutoAnalyzer*, we optimized two parallel programs with performance improvements by minimally 20% and maximally 170%.

In near future, we will extend our method to more generalized parallel applications beyond SPMD style.